\documentclass[aps,prl,showpacs,twocolumn,floats]{revtex4}
\usepackage{amssymb}

\usepackage{graphicx}
\usepackage{bm}

\begin{document}

\bibliographystyle{prsty}
\author{Eugene M. Chudnovsky$^{1,2}$, Javier Tejada$^1$, Ricardo Zarzuela$^1$}
\affiliation{$^{1}$Facultat de F\'{i}sica, Universitat de
Barcelona, Diagonal 645, 08028 Barcelona, Spain\\
$^{2}$Physics Department, Lehman College, The City University of New
York, 250 Bedford Park Boulevard West, Bronx, NY 10468-1589}
\date{\today}

\begin{abstract}
Two-state systems may exhibit mechanical forces of purely quantum
origin that have no counterpart in classical physics. We show that
the such forces must exist in molecular magnets due to quantum
tunneling between classically degenerate magnetic states. They can
be observed in the presence of a microwave field when the magnet is
placed in a static magnetic field with a gradient.

\end{abstract}

\title{Quantum Forces in Molecular Magnets}

\maketitle

Forces of quantum nature are known in physics. One example is a
Casimir force between two surfaces in a close proximity to each
other. It is caused by the quantization of the fields in the space
between the surfaces \cite{Casimir}. Casimir forces have been
extensively discussed in relation to a number of condensed matter
systems \cite{Kardar}, see, e.g., recent applications to topological
insulators \cite{Grushin-PRL2011}. In this Letter we will discuss
the force of a purely quantum origin of another kind: The force that
is pertinent to the two-state systems. Such systems are very common
in nature. They correspond to a situation when the lowest energy
doublet of a quantum system is separated from the rest of the
spectrum by a large gap, making only that doublet relevant in
low-energy experiments. We will focus our attention on molecular
magnets, although our conclusions will apply to any two-state system
for which the energy distance $\Delta$ between the two states of the
doublet can be controlled by the external field. Quantum Hamiltonian
of a non-interacting two-state system is
\begin{equation}\label{H}
{\cal{H}} = -\frac{1}{2}\Delta \sigma_z
\end{equation}
where $\sigma_z$ is a Pauli matrix. Its eigenstates, $| \pm
\rangle$, correspond to $\sigma_z = \pm 1$ and have energies
$E_{\pm} = \mp \Delta/2$, with $| + \rangle$ being the ground state
and $| - \rangle$ being the excited state. The general form of the
normalized wave function is a superposition of the $| \pm \rangle$
states:
\begin{equation}\label{Psi}
| \Psi \rangle = C_+ e^{-iE_+ t/\hbar}| + \rangle + C_- e^{-iE_-
t/\hbar}| - \rangle
\end{equation}
with $|C_{+}|^2 + |C_{-}|^2 = 1$.

The Hamiltonian (\ref{H}) is equivalent to the Hamiltonian of a
spin-$1/2$ particle in the magnetic field. In the presence of the
field gradient, there is a force on the particle that was used at
the dawn of quantum physics to separate particles in beams according
to their spin projection \cite{SG}. The effect we are after has the
same origin. However, here we are particularly interested in the
situation when $\sigma_z$ describing the two-state system has
nothing to do with the real spin $1/2$, and the magnetic moment
associated with it, but is rather related to the tunnel splitting of
classically degenerate states.

Let us consider a crystal containing a macroscopic number of
non-interacting two-state particles. The occupation numbers of the
states with $\sigma_z = \pm 1$ are $n_{\pm} = |C_{\pm}|^2$. The
corresponding one-particle density matrix operator is $\rho =
|\Psi(t)\rangle\langle \Psi(t)|$. In the presence of the gradient of
$\Delta$ created by the gradient of the external field, the force on
the crystal is
\begin{equation}\label{F}
{\bf F} = \sum \,Tr\left[\left(-{\bm \nabla}{\cal{H}}\right)
\rho\right] = \frac{1}{2}\sum(n_+ - n_-) ({\bm \nabla} \Delta)
\end{equation}
where summation is over occupation numbers of the particles. Note
that ${\bf F}$ depends only on the gradient of $\Delta$ and on the
occupation numbers, but not on the choice of the quantization axis
for the effective spin $1/2$.

The physical origin of the above force is clear. The particles with
$\sigma_z = 1$, occupying the ground state level with energy $E_+ =
-\Delta/2$, are attracted to the area where $\Delta$ is higher,
while the particles with $\sigma_z = -1$, occupying the excited
level with energy $E_- = \Delta/2$ are attracted to the area where
$\Delta$ is lower. If for all particles $\Delta \ll k_B T$ the two
states are almost equally occupied, $n_+ \approx n_-$, and the force
is close to zero but at $k_B T \ll \Delta$ one has $n_+ \approx 1,
n_- \approx 0$, and the net force is non-zero.

In molecular magnets this force must exist as a consequence of
quantum tunneling between classically degenerate magnetic states
\cite{book}. According to Eq.\ (\ref{F}) at $k_BT \ll \Delta$ (that
is, at $n_+ \approx 1$, $n_- \approx 0$) and the gradient of the
tunnel splitting $|{\bm \nabla} \Delta| \sim 1$K/cm, the force on a
small crystal containing $10^{16}$ magnetic molecules would be of
order $10^{-5}$N. On increasing temperature above $\Delta/k_B$ the
force must disappear due to the equilibration of the populations of
the two levels.

Another interesting situation is when the populations $n_{\pm}$
oscillate in time, causing mechanical oscillations of the crystal.
This could be measured with the help of a mechanical resonator, a
force microscope or a tunneling microscope. Such a situation can be
achieved by placing the crystal in the ac field that provides Rabi
oscillations of the populations of the two levels \cite{Rabi}. Some
complication comes from the gradient of $\Delta$ that makes it
impossible to satisfy the condition of the resonance, $\hbar \omega
= \Delta$, by the ac field of a fixed frequency $\omega$, for all
particles of the crystal. Below we compute the oscillating force on
the crystal of a molecular magnet and study conditions under which
it can be experimentally detected.

The exact form of the spin Hamiltonian is not important for our
purpose but to explain the concept we will stick for the moment to a
toy model of a crystal of uniaxial high-spin magnetic molecules in a
strong transverse magnetic field. We choose the easy magnetization
axis in the $Y$-direction. The field, $B(z)$, is applied in the
$X$-direction, with the field gradient being in the $Z$-direction.
The Hamiltonian of an individual molecule is
\begin{equation}
{\cal{H}}_M = -DS_y^2 -g \mu_B B S_x
\end{equation}
where $D$ is the anisotropy constant, $g$ is the gyromagnetic
factor, and $\mu_B$ is the Bohr magneton. Non-commutation of ${{\cal
H}}_M$ with $S_y$ provides tunneling between degenerate classical
energy minima, $S_y = \pm S$ \cite{Mn12,book}. When $B$ is small
compared to the anisotropy field $2DS/(g\mu_B)$ the tunnel splitting
can be computed with the help of the perturbation theory
\cite{Garanin,lectures}
\begin{equation}\label{Delta}
\Delta = \frac{8DS^2}{(2S)!}\left(\frac{g\mu_B B}{2D}\right)^{2S}
\end{equation}
At higher fields $\Delta$ can be obtained by exact diagonalization
of the spin Hamiltonian. In the last 20 years this has been done for
a number of molecular magnets with account of all terms in the
Hamiltonian that are dictated by symmetry. High-field cavity and EPR
experiments
\cite{Barco-PRB2000,Luis-PRL2000,Takahashi-PRB2004,Takahashi-Nature2011}
in spin-10 Fe-8 and Mn-12 molecular magnets agree well with
theoretical values of $\Delta$. Strong dependence of $\Delta$ on $B$
makes it relatively easy to create a significant gradient of
$\Delta$ inside the sample.

Note that at the classical level the magnetic field applied
perpendicular to the easy magnetization axis of the crystal creates
a finite magnetic moment ${\bf M}$ in the direction of the field.
The gradient of the field then creates a constant classical force on
the paramagnet, ${\bf F} = {\bm \nabla} ({\bf M} \cdot {\bf B})$.
Unlike this force, the quantum force we are after appears due to the
tunnel splitting $\Delta$. To measure this force one should exploit
its dependence on the populations, $n_{\pm}$, of the tunneling
doublet. With an eye on resonant experiments we are interested in
the situation where the change in $\Delta$ across the sample of
thickness $a$ is small compared to $\Delta$ itself, that is,
$a|\nabla \Delta| \ll \Delta_0$, where $\Delta_0$ is the tunnel
splitting in the middle of the sample. According to Eq.\
(\ref{Delta}) this translates into the condition $Sa|dB/dz| \ll
B_0$, where $B_0$ is the transverse field in the middle of the
sample. Writing $B = B_0(1+z/l)$, where $l$ is the characteristic
length describing the field gradient, the above conditions can be
reduced to $a \ll l/S$. The dependence of $\Delta$ on $z$ is then
given by
\begin{equation}
\Delta(z) = \Delta_0\left(1+ \frac{2Sz}{l}\right), \;\;\; -a/2 < z <
a/2
\end{equation}

Let now a weak ac field of amplitude $h_{ac}$ and frequency $\omega
=\Delta_0/\hbar$ be applied along the $Y$-axis. Classically, such a
field does not generate any magnetic moment and, therefore, it
cannot be responsible for any classical force. In the presence of
the ac field the truncated two-state Hamiltonian for an individual
molecule becomes
\begin{equation}\label{H-ac}
{\cal{H}} = -\frac{1}{2}\Delta \sigma_z - W\sigma_y \cos(\omega t)
\end{equation}
where
\begin{equation}
W = g\mu_B S h_{ac}
\end{equation}
For the moment we will focus on the detuning from the resonance
solely due to the field gradient,
\begin{equation}\label{delta-z}
\delta(z) = \frac{\Delta(z)}{\hbar} - \omega = \frac{2Sz}{l}\omega
\end{equation}
and will comment on the detuning due to dipolar and hyperfine fields
later on.

Close to the resonance, solution of the Schr\"{o}dinger equation
with the Hamiltonian (\ref{H-ac}) and the wave function (\ref{Psi})
permits the rotating wave approximation, yielding the famous Rabi
result \cite{Rabi}
\begin{eqnarray}\label{C}
C_- & = & i\frac{W}{\hbar\Omega_R}e^{i\delta
t/2}\sin\left(\frac{\Omega_R t}{2}\right) \nonumber \\
C_+ & = & e^{-i\delta t/2}\left[\cos\left(\frac{\Omega_R
t}{2}\right)+ i\frac{\delta}{\Omega_R}\sin\left(\frac{\Omega_R
t}{2}\right)\right]
\end{eqnarray}
where
\begin{equation}\label{Rabi-f}
\Omega_R(z) = \sqrt{\delta^2(z) + \left(\frac{W}{\hbar}\right)^2}
\end{equation}

As has been discussed above, at $T = 0$ the force acting on the
molecules is proportional to
\begin{equation}\label{n-pm}
n_+ - n_-  =  |C_+|^2 - |C_-|^2 \nonumber \\
=  \frac{\delta^2}{\Omega_R^2} + \frac{W^2}{(\hbar \Omega_R)^2}
\cos(\Omega_R t)
\end{equation}
which is now a function of coordinates due to the field gradient.
Switching from summation to integration in Eq.\ (\ref{F}), we have
\begin{equation}\label{Force}
{\bf F}(t) = \frac{1}{2}N({\bm
\nabla}\Delta)\int_{-a/2}^{a/2}\frac{dz}{a}\left[|C_+(z,t)|^2 -
|C_-(z,t)|^2\right]
\end{equation}
where $N$ is the number of molecules in the crystal. When $W
\rightarrow 0$, one has $n_+ - n_- \rightarrow 1$ and the problem
reduces to the one of a constant force. Here we are interested in
the oscillating part of the force due to the second term in Eq.\
(\ref{n-pm}),
\begin{equation}\label{F-ac}
F_{osc} = NS \frac
{\Delta_0}{l}\int_{-a/2}^{a/2}\frac{dz}{a}\frac{W^2}{[\hbar
\Omega_R(z)]^2}\cos\left[\Omega_R(z)t\right]
\end{equation}

We shall introduce notation
\begin{equation}\label{omega-0}
\Omega_0 = \Omega_R(0) = \frac{W}{\hbar} = S\left(\frac{g\mu_B
h_{ac}}{\hbar}\right)
\end{equation}
for the Rabi frequency at $z=0$. The condition $\delta(z=a/2) \ll
\Omega_0$ is needed to insure that contributions of molecules
belonging to different layers of the crystal do not cancel for a
substantial period of time. It is equivalent to
\begin{equation}\label{condition}
\frac{Sa}{l}\omega \ll \Omega_0
\end{equation}
With this condition Eq.\ (\ref{F-ac}) becomes
\begin{equation}\label{F-epsilon}
F_{osc} = N \frac{\hbar \Omega_0}{a}\int_0^{\epsilon}d\xi
\frac{\cos\left(\Omega_0 t\sqrt{1+\xi^2}\right)}{1+\xi^2}
\end{equation}
where
\begin{equation}\label{epsilon}
\epsilon =
\left(\frac{Sa}{l}\right)\left(\frac{\omega}{\Omega_0}\right) \ll 1
\end{equation}

Smallness of $\epsilon$ allows one to replace $\sqrt{1 + \xi^2}$
with $1+\xi^2/2$ and to reduce the integral in Eq.\
(\ref{F-epsilon}) to the sum of Fresnel integrals
\begin{equation}\label{expression}
\cos(\Omega_0 t)\int_0^{\epsilon}d\xi\cos\left(\frac{\Omega_0 t
\xi^2}{2}\right) -\sin(\Omega_0
t)\int_0^{\epsilon}d\xi\sin\left(\frac{\Omega_0 t \xi^2}{2}\right)
\end{equation}
There are two regimes in the temporal behavior of this expression.
At $\Omega_0 t \ll 2/\epsilon^2$ it reduces to
$\epsilon\cos(\Omega_0 t)$, resulting in
\begin{equation}\label{force-1}
F_{osc} = NS \frac{\Delta_0}{l}\cos(\Omega_0 t)
\end{equation}
At $\Omega_0 t \gg 2/\epsilon^2$, recalling that $\int_{0}^{\infty}
dx \cos x^2 = \int_{0}^{\infty} dx \sin x^2 = \sqrt{\pi/8}$, one
obtains
\begin{equation}\label{force-2}
F_{osc} = N \frac{\hbar \Omega_0}{a} \sqrt{\frac{\pi}{2\Omega_0
t}}\cos\left(\Omega_0 t + \frac{\pi}{4}\right)
\end{equation}
that is, the force that goes down as a square root of time.
Interpretation of the two regimes is straightforward. Initially all
molecules oscillate in phase due to the small variation in the Rabi
frequency. With time, however, molecules belonging to different
layers of the crystal accumulate large phase differences and the
forces acting on them begin to cancel.

\begin{figure}[!ht]
\vspace{-0.5cm}
\includegraphics[width=94mm]{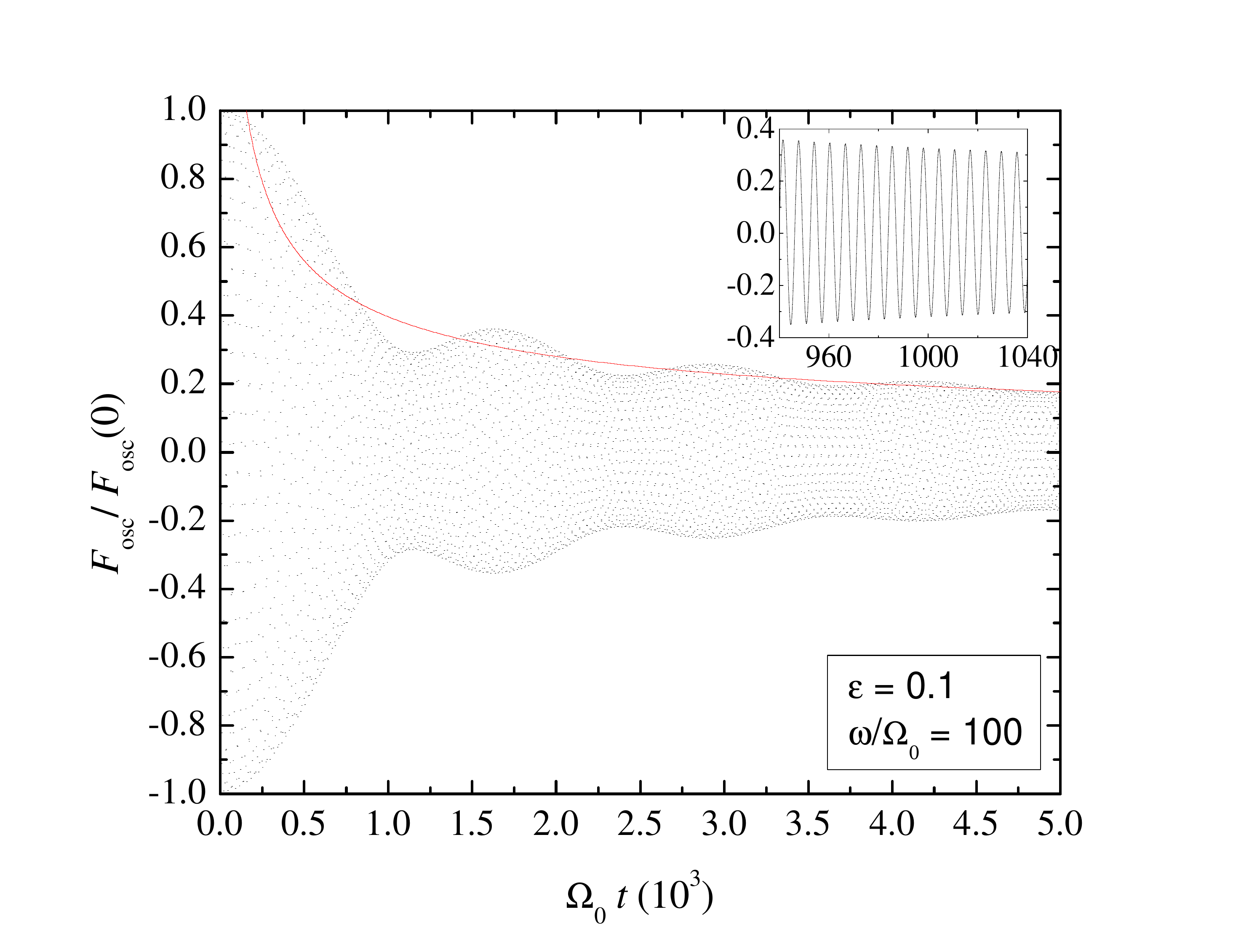}
\vspace{-0.5cm} \caption{Oscillating force normalized by its value
at $t=0$ for $\epsilon = 0.1$ and $\omega/\Omega_0 = 100$. The inset
shows oscillations. The envelope line shows analytical result for
the amplitude of the force at long times.} \label{fig-force1}
\end{figure}

To solve the problem for an arbitrary microwave pulse and arbitrary
$\epsilon$, without relying on the rotating wave approximation, we
replace $W$ in Eq.\ (\ref{H-ac}) with $Wf(t)$, where $W$ is the same
constant as before and $f(t)$ is an arbitrary function of time
representing the time dependence of the amplitude of the ac field.
Schr\"{o}dinger equation generates the following equations for
$C_{\pm}$:
\begin{equation}\label{eq-C}
i\frac{d}{d\tau} C_{\pm}(\bar{z},\tau) = \frac{1}{2}e^{\mp
2i\epsilon \bar{z}
\tau}f(\tau)\left[1+e^{{\mp}2i(\omega/\Omega_0)\tau}\right]C_{\mp}(\bar{z},\tau)
\end{equation}
where $\tau \equiv \Omega_0 t$ and $\bar{z} \equiv z/a$. The
solution depends on two parameters: $\epsilon$ and
${\omega}/\Omega_0$. We solve Eqs.\ (\ref{eq-C}) numerically and
compute the force given by Eq.\ (\ref{Force}). The oscillating part
of the force for $f(t) = 0$ at $t < 0$ and $f(t) = 1$ at $t > 0$, in
the case of $\epsilon = 0.1$ and $\omega/\Omega_0 = 100$, is shown
in Fig. \ref{fig-force1}.
For small $\epsilon$ the deviation of numerical results from the
analytical formulas (\ref{force-1}) and (\ref{force-2}), obtained
for the two limiting cases of short and long times, is small. The
envelope curve in Fig. \ref{fig-force1} follows the asymptotic
analytical result at long times, $F_{osc}(t)/F_{osc}(0) =
(1/\epsilon)\sqrt{\pi/(2\Omega_0 t)}$.

\begin{figure}[!ht]
\vspace{-0.5cm}
\includegraphics[width=94mm]{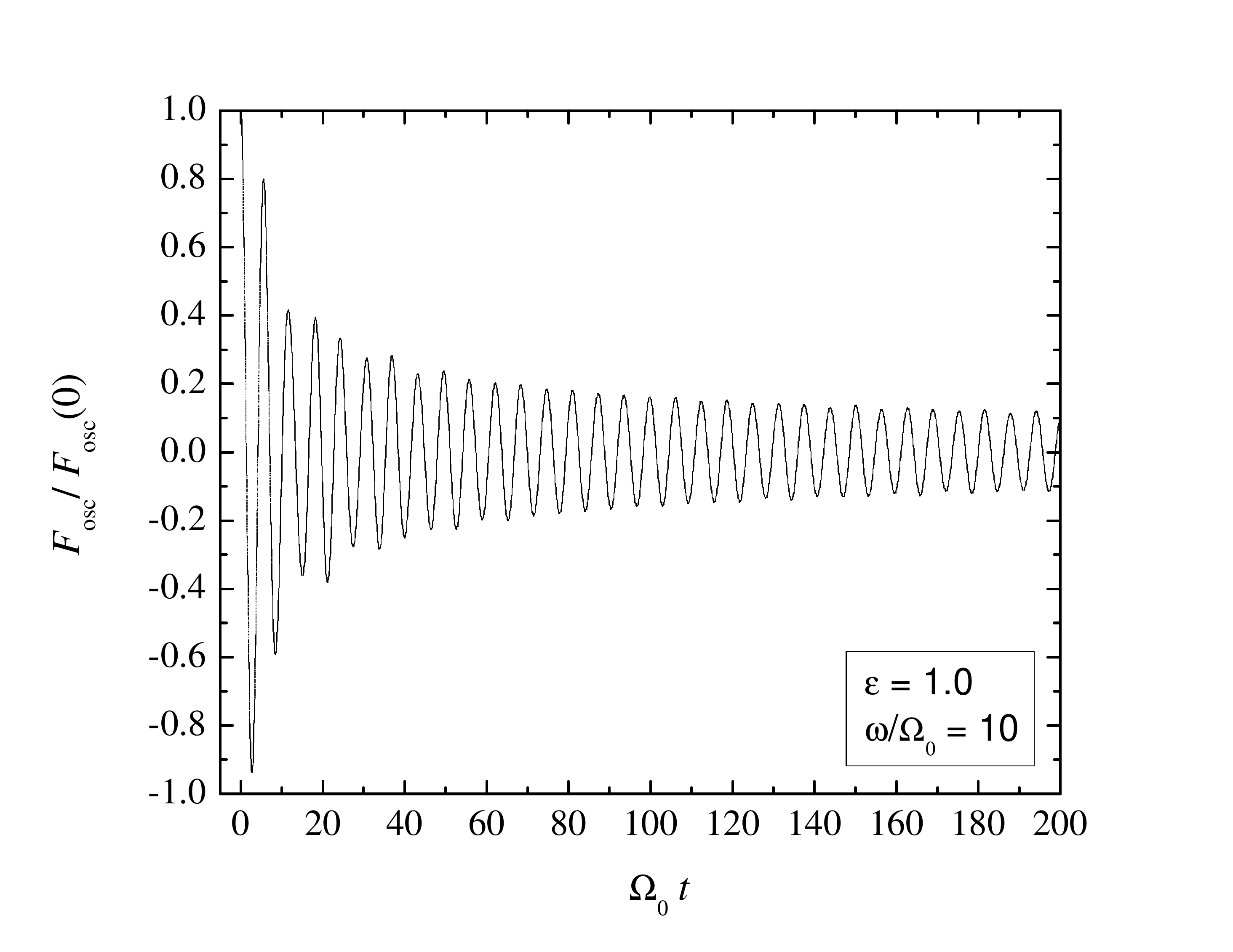}
\vspace{-0.5cm} \caption{Oscillating force normalized by its value
at $t=0$ for $\epsilon = 1$ and $\omega/\Omega_0 = 10$.}
\label{fig-force2}
\end{figure}
Note that a sizable force may also exist in the case of a strong
detuning due to a strong field gradient, when the condition
$\epsilon \ll 1$ is not fulfilled. For instance, the ratio of
$F_{osc}(0)$ at $\epsilon = 1$ to $F_{osc}(0)$ at $\epsilon = 0.1$
is $0.788$ and the decrease of the amplitude with time is rather
slow. This occurs due to the coherent contribution to the force of
the central layer of the crystal where the detuning is still small.
Fig. \ref{fig-force2} shows the time dependence of the normalized
oscillating force for $\epsilon = 1$ and $\omega/\Omega_0 = 10$.

We shall now focus on the conditions needed to observe the
oscillating force. Most of these conditions are the same as the ones
needed for detecting Rabi oscillations. Firstly, the spread of
$\Delta$ due to dipole-dipole and hyperfine interactions should be
small to provide near-resonance condition for all molecules apart
from broadening due to the gradient of the transverse field. Local
stray field $H_L$ along the anisotropy axis would change $\Delta_0$
to $\Delta = \sqrt{\Delta_0^2 + W_L^2}$ where $W_L=g S \mu_B H_L$.
Small spread of $\Delta$ occurs when $W_L \ll \Delta_0$, in which
case $\Delta = \Delta_0[1+W_L^2/(2\Delta_0)]$. For $H_L \sim 10^2$G
small spread of $\Delta$  requires $\omega=\Delta_0/\hbar$ in the
GHz range. Some molecular magnets have $\Delta$ in this range or
higher at $B = 0$ due to the symmetry of the Hamiltonian
\cite{Waldman-PRL2005,Loubens-JAP2008}. For the transverse field to
dominate $\Delta$ of that magnitude the field must be in the tesla
range. For $B_0 = 1$T and the field gradient of $1$T/cm (which can
be easily achieved in a quadrupole magnet) the parameter
$l=B_0/|\nabla B|$ is $1$cm.

Secondly, a significant number of Rabi cycles should occur before
the phase of the wave function of an individual molecule is
destroyed by decoherence. Studies of decoherence
\cite{Hill-Science2003,Barco-PRL2004,Wernsdorfer-PRB2005,Ardavan-PRL2007,Loubens-JAP2008,Bertaina-Nature,Schledel-PRL2008,Takahashi-Nature2011}
suggest that the decoherence time in molecular magnets can hardly
exceed one hundred nanosecond. Thus, the Rabi frequencies involved
must be in the excess of $10$MHz, which requires $h_{ac}$ of a few
gauss or greater. To date Rabi oscillations have been observed in
the $S = 5$ Fe-5 molecular magnet \cite{Schledel-PRL2008} and in the
$S = 1/2$ V-15 molecular magnet \cite{Bertaina-Nature,Yang-PRL2012}.

For $\omega \sim 10$GHz and $\Omega_0 \sim 100$MHz, the condition
(\ref{epsilon}) at $l \sim 1$cm and $S \sim 1$, gives $a \ll 0.1$mm.
Molecular magnet of such dimensions would typically have less than
$10^{15}$ molecules. Substituting $N = 10^{14}$, $\omega = 10$GHz,
$l = 1$cm, $S = 1$ into Eq.\ (\ref{force-1}), we get the amplitude
of the oscillating force of order $10^{-8}$N, which is comparable to
the weight of the crystal. The force of that magnitude should be
powerful enough to generate surface acoustic waves in the substrate
to which the sample is fixed. It could also be measured by placing
the sample on a microcantilever and measuring the deflection of the
cantilever as a function of temperature. Since the Rabi frequency is
proportional to the amplitude of the ac field, the resonance with
the mechanical mode of the cantilever can be achieved by varying
$h_{ac}$. Given the magnitude of the computed force the effect may
be detectable even when only a small fraction of the molecules is
near resonance with the ac field due to a large field gradient or
because of dipolar, hyperfine, and other stray fields in the
crystal. This may also be true for a system that is artificially
diluted to reduce dipole-dipole interactions.

In conclusion, we have shown that two-state systems can exhibit
forces of purely quantum origin. Forces related to quantum tunneling
of the magnetic moment have been computed. High magnitude of such
forces and their strong dependence on temperature and field gradient
should make them detectable in molecular magnets.

The work of EMC has been supported by the University of Barcelona
and by the NSF Grant No. DMR-1161571. The work of JT and RZ has been
supported by the Spanish Government Project No. MAT2011- 23698.

\end{document}